\documentclass[preprint,12pt]{elsarticle}
\usepackage{amssymb}
 \usepackage{amsmath}
\usepackage{color}

\begin{document}

\begin{frontmatter}

\title{Optimal values of bipartite entanglement in a tripartite system}

\author{Shaon Sahoo}

\address{Department of Physics, Indian Institute of Science, Bangalore 560012, India \\
Solid State and Structural Chemistry Unit, Indian Institute of Science,
Bangalore 560012, India}

\begin{abstract}
For a general tripartite system in some pure state, an observer possessing any two 
parts will see them in a mixed state. By the 
consequence of Hughston-Jozsa-Wootters theorem, each 
basis set of local measurement on the third part will correspond to a particular 
decomposition of the bipartite mixed state into a weighted sum of pure states.
It is possible to associate an average bipartite entanglement ($\bar{\mathcal{S}}$) 
with each of these decompositions. The maximum value of $\bar{\mathcal{S}}$ is 
called the entanglement of assistance ($E_A$) while the minimum value is called the 
entanglement of formation ($E_F$). An appropriate choice of the basis set of local 
measurement will correspond to an optimal value of $\bar{\mathcal{S}}$; we find here 
a generic optimality condition for the choice of the basis set. In the present 
context, we analyze the tripartite states $W$ and $GHZ$ and show how they are 
fundamentally different.
\end{abstract}


\end{frontmatter}

\section{Introduction}
\noindent
Besides its fundamental importance in interpreting and understanding quantum 
mechanics, the quantum entanglement has attracted an immense interest in recent 
times because of its potential to play a significant role in modern technology. 
In addition, the quantum entanglement has now become a 	powerful tool to study 
the quantum many-body systems \cite{amico08}. To be able to use the entanglement 
effectively and efficiently, it is necessary to characterize and quantify it 
by meaningful ways. There are many entanglement measures proposed for this 
purpose. 
In the axiomatic approach, there are some conditions to be satisfied for an 
entanglement measure to be a proper monotone \cite{horodecki09,vidal00}. For 
pure bipartite state, the {\it von Neumann entropy} is a suitable and widely 
used entanglement measure. Unfortunately it is not a good measure for non-pure or 
mixed bipartite state. For the mixed states, two popular entanglement measures 
are the {\it concurrence} \cite{wootters97} and {\it negativity} \cite{vidal02}. 
These two measures are reliable for distinguishing entangled states from 
separable states respectively for $2\times2$ system, and $2\times2$ and 
$2\times3$ systems. Generalization of these measures to higher dimensions is 
possible, but most of the time they are not satisfactory and unique 
\cite{horodecki09,plenio07,mintert05}. 

The entanglement of formation ($E_F^{\infty}$) \cite{bennet96} for a mixed 
state is a more
general concept than the concurrence. For $2\times2$ system, the concurrence 
is monotonically related to $E_F^{\infty}$ \cite{wootters01}. 
Even though $E_F^{\infty}$ is a good measure for any bipartite mixed state, 
it is extremely difficult to calculate. The reason can be understood from 
the definition of the quantity, as shown below,
\begin{eqnarray}
\label{entfrm}
E_F^{\infty} (\rho)= {\rm inf} \left\{\sum_i p_i \mathcal{S}_i
(|\psi_i\rangle) ~:~  p_i \geq 0, \sum_i p_i = 1;~ 
\rho = \sum_i p_i |\psi_i\rangle \langle \psi_i| \right\}.
\end{eqnarray}
Here $\mathcal{S}_i$ can be any good entanglement measure for the pure 
bipartite state $|\psi_i\rangle$ (e.g. it can be von Neumann entropy). 
Basically, to find $E_F^{\infty}$ one has to do 
minimization of an average quantity over infinite possible decomposition 
of the given mixed state ($\rho$).

The entanglement of assistance ($E_A^{\infty}$) is another measure for an 
arbitrary bipartite mixed state \cite{vincenzo98}. 
Though it is not a proper monotone 
\cite{gour06}, it got importance due to 
its possible application in quantum technology. It is defined as the maximum 
possible average entropy between two parties, as shown below,
\begin{eqnarray}
\label{entasst}
E_A^{\infty} (\rho)= {\rm sup} \left\{\sum_i p_i \mathcal{S}_i
(|\psi_i\rangle) ~:~  p_i \geq 0, \sum_i p_i = 1;~ 
\rho = \sum_i p_i |\psi_i\rangle \langle \psi_i| \right\}.
\end{eqnarray}
 
Quantifying and classifying multipartite entanglement is a very hard problem; 
it is now generally accepted that a single number is not enough for the 
purpose \cite{horodecki09}. In this work we concentrate on studying a general 
tripartite system in a pure state. 
First we note that, according to the Hughston-Jozsa-Wootters (HJW) theorem 
\cite{hughston93}, any finite decomposition compatible with a given mixed 
bipartite state can be created by a local measurement on a third part added 
in a pure state with the bipartite system. For a general tripartite system 
($A-B-C$) in pure state, 
this theorem implies that, any local measurement on a part ($C$) will 
correspond to a particular decomposition of the bipartite mixed state that the 
other two parts ($A-B$) are effectively in. 
It may be here worth mentioning that all the 
quantum measurements in this work are
considered to be non-selective projective-type (von Neumann measurement).  
One can associate an average entanglement 
($\bar{\mathcal{S}}$) with every decomposition of a mixed state; 
very often this quantity is called the entanglement `localized' between two 
parts of a tripartite system by doing a local measurement on the third part. 
This average quantity is always positive and have a upper 
bound for a finite dimensional system. This implies that $\bar{\mathcal{S}}$ 
will have optimal (maximum/minimum) values and as a consequence of the HJW 
theorem, these optimal values will correspond to some basis sets of 
measurement. In other words, since the decomposition of mixed state changes 
with the measurement basis, it is possible to get optimal values of the 
quantity $\bar{\mathcal{S}}$ by choosing appropriate measurement basis sets.
The maximum and minimum values of $\bar{\mathcal{S}}$ are termed respectively 
as the entanglement of assistance ($E_A$) and the entanglement of formation 
($E_F$) for the given tripartite pure state ($|\Psi\rangle$). 
These two quantities are given by following equations,
\begin{eqnarray}
\label{entfrm1}
E_F (|\Psi\rangle)= {\rm inf} \left\{\sum_i p_i \mathcal{S}_i
(|\psi_i\rangle^{AB}) ~:~  p_i \geq 0, \sum_i p_i = 1; \right. \\ \nonumber 
\left. |\Psi\rangle = \sum_i \sqrt{p_i} 
|\phi_i \rangle^C |\psi_i\rangle^{AB} \right\}. \\ 
\label{entasst1}
E_A (|\Psi\rangle)= {\rm sup} \left\{\sum_i p_i \mathcal{S}_i
(|\psi_i\rangle^{AB}) ~:~  p_i \geq 0, \sum_i p_i = 1; \right. \\ \nonumber 
\left. |\Psi\rangle = \sum_i \sqrt{p_i} 
|\phi_i \rangle^C |\psi_i\rangle^{AB} \right\}.
\end{eqnarray}
Here $|\phi_i \rangle^C$s form an orthonormal basis set of $C$ and this set is 
used as a basis set for local measurement. After a measurement, parts $A$ and $B$ 
jointly assume pure state $|\psi_i\rangle^{AB}$ with probability $p_i$. In general 
$|\psi_i\rangle^{AB}$s are not orthogonal to each other (see Section 2). 

Here it may be briefly mentioned that though there are some 
attempts to generalize the definition of the entanglement of formation 
to the multipartite systems (see for example \cite{wang00}), but proper 
generalization is not possible until we have a clear notion of maximally 
entangled multipartite states \cite{plenio07,plenio01}. This notion is still 
lacking; situation here is even more complicated due to the existence of 
different classes of multipartite states (for example, there 
are two non-interconvertible classes of tripartite states \cite{dur00}).
The definitions of the entanglement of formation and the 
entanglement of assistance given in this work for the pure tripartite states 
(cf. Eqs. (\ref{entfrm1}) and (\ref{entasst1})) are not an attempt to define 
any tripartite entanglement monotones by generalizing the corresponding 
bipartite monotones; these definitions would be though very useful and 
will serve us two 
purposes. Besides giving some important informations about the
given tripartite state (see Section 3), they will help us calculate 
$E_A^{\infty}$ and $E_A^{\infty}$ for a given bipartite 
mixed state by using the concept of {\it ancilla} (as deliberated below).
 
It is
known that, a person possessing two parts ($A$ and $B$) of a tripartite system 
(which is in a pure state $|\Psi\rangle$) will only see a reduced 
state $\rho^{AB} =  
{\rm Tr}_C (|\Psi\rangle \langle \Psi|)$. As a consequence of the 
HJW theorem, each basis set of measurement of $C$ 
corresponds to a particular decomposition of the mixed state $\rho^{AB}$. 
The number of pure states appearing in a decomposition cannot exceed the basis 
set dimension of $C$ (say, $D_C$) \cite{hughston93}. On the other hand, the number 
of terms in an unrestricted decomposition of a given
mixed bipartite state (without 
reference to $C$ or any pure tripartite state) can be in principle any large number. 
According to the variational principle, the unrestricted minimization 
of the average quantity $\bar{\mathcal{S}}$ will give a lower value than the 
the restricted minimization of the quantity. This implies that, the 
entanglement of formation ($E_F$) of a pure tripartite state
as defined in Eq. (\ref{entfrm1}) is higher than 
$E_F^{\infty}$ for the corresponding reduced state ($\rho^{AB}$), i.e., 
$E_F^{\infty}(\rho^{AB}) \leq E_F(|\Psi\rangle)$. Following a similar line of 
argument we can also say that $E_A^{\infty}(\rho^{AB}) \geq E_A(|\Psi\rangle)$.

In general finding $E_F^{\infty}$ or $E_A^{\infty}$ for an arbitrary bipartite 
mixed state is difficult. As we mentioned before, in principle there can be any 
number of terms in an unrestricted decomposition of a mixed state. This makes 
even numerical calculations of the quantities really hard. It is though 
speculated that, to evaluate
$E_F^{\infty}$ or $E_A^{\infty}$ for a mixed bipartite state $\rho$, it 
is enough to consider only a finite number of terms in the decomposition of 
$\rho$. For example, it was proved that it is sufficient to consider only $r^2$ 
terms in a decomposition to find $E_F^{\infty}$, where $r$ is the rank of the 
mixed state \cite{uhlmann97}. In fact it turned out that, for $2\times2$ 
systems it is enough to consider only four states for the purpose 
\cite{sanpera98,wootters98}. Therefore to find 
$E_F^{\infty}$ or $E_A^{\infty}$ 
for a given mixed state, we can start with a pure tripartite state for which 
the reduced state $\rho^{AB}$ is the same as the given mixed state. The value
of $E_F$ ($E_A$) obtained from the tripartite state would not be the same as 
$E_F^{\infty}$ ($E_A^{\infty}$) if the number of terms required 
for the optimization is more than $D_C$ (basis set 
dimension of $C$). To increase the basis set dimension, 
one can add a suitable {\it ancilla} to the the third part 
$C$ and do a joint measurement \cite{hughston93}.

Since the optimization process to find $E_F$ or $E_A$ is very demanding, 
to help doing that, we derive in this paper an {\it optimality condition}. 
The basis set of measurement which satisfies this condition will correspond 
to an optimal value of the average entropy $\bar{\mathcal{S}}$.

In the last part of this paper, we analyze two tripartite states $W$ and $GHZ$, 
and show how they are fundamentally different in the present context.

\section{The optimality condition}
\noindent
Let $A$, $B$ and $C$ be three parts of a tripartite system in the pure state 
$|\Psi\rangle$. A local quantum measurement on $C$ by some basis set would 
result in $S$ (= $A$ + $B$) assuming different pure states with appropriate 
probabilities.

When expressed in the product basis states of $C$ and $S$, the given tripartite 
state becomes, 
\begin{eqnarray}
\label{trprt}
|\Psi\rangle = \sum_{i,j=1,1}^{D_C,D_S}g_{i,j} 
|\xi_i\rangle^C |\phi_j\rangle^S. 
\end{eqnarray}
Here $|\xi_i\rangle^C$s
($|\phi_i\rangle^S$s) are some orthonormal basis vectors of the state space of
$C$ ($S$) with dimensionality $D_C$ ($D_S$). This state is assumed to be 
normalized: $\sum_{i,j=1,1}^{D_C,D_S}g_{i,j}g_{i,j}^* = 1$.
Let us now rewrite this state in the following special form, 
\begin{eqnarray}
\label{wvfn}
|\Psi\rangle = \sum\nolimits_{i=1}^{D}
\sqrt{p_i}|\xi_i\rangle^C |\xi_i\rangle^S, 
\end{eqnarray}
with $p_i = \sum_{j'=1}^{D_S}g_{i,j'}g_{i,j'}^*$ and $|\xi_i\rangle^S =
\sum_{j=1}^{D_S}\frac{g_{i,j}}{\sqrt{p_i}}
|\phi_j\rangle^S$. Here the summation runs over nonzero $p_i$'s, numbering
$D$ ($\le D_C$). In general, states $|\xi_i\rangle^S$s are not orthogonal 
(but they all are normalized).
The operational interpretation of the later expression of the state
$|\Psi\rangle$ given in Eq.~(\ref{wvfn}) is that, if we perform a local 
quantum measurement on $C$ by the basis set $\{\xi^C\}$, the state 
$|\Psi\rangle$ will 
collapse and we will get $S$ in different pure states $|\xi_i\rangle^S$s  
with corresponding probabilities $p_i$'s.

If $\mathcal{S}_i$ is the von Neumann entropy of the pure bipartite state 
$|\xi_i\rangle^S$, then after the measurement, the
average entropy (quantifying average entanglement) localized between $A$ 
and $B$ would be,
\begin{eqnarray}
\label{aventr}
\bar{\mathcal{S}}\{\xi^C\} = \sum\nolimits_{i=1}^{D} 
p_i \mathcal{S}_i.
\end{eqnarray}
As both $p_i$'s and $\mathcal{S}_i$'s depend on the choice of the basis 
set of measurement, $\{\xi^C\}$, the average entropy $\bar{\mathcal{S}}$ 
will also depend on the choice of the basis set. We will now derive a 
condition for the choice of the basis set which optimizes 
$\bar{\mathcal{S}}$.

We first note that, any basis set of measurement can be obtained from 
an (arbitrary) initial basis set $\{\xi^C\}$ 
by application of a series of {\it elementary transformation}s (ETs).
Here an ET is a small-angle orthonormal transformation (rotation) of any 
two basis states keeping others unchanged. We now derive first
order change in $\bar{\mathcal{S}}$ due to an ET. If $|\xi_i\rangle^C$ and
$|\xi_j\rangle^C$ are any two initial basis states, then the two new basis 
states obtained by an ET would be,
\begin{eqnarray}
\label{newbasis}
|\xi'_i \rangle^C =|\xi_i\rangle^C +\epsilon |\xi_j\rangle^C ~{\rm and}~
|\xi'_j \rangle^C =|\xi_j\rangle^C -\epsilon |\xi_i\rangle^C.
\end{eqnarray}
Here $\epsilon$ is the small angle (a parameter) whose higher order
terms can be neglected. We may note that these two new basis states 
are orthogonal (i.e., $^C\langle \xi'_i|\xi'_j \rangle^C = 0$) and 
normalized within first order approximation.
Due to change in the basis states of measurement, corresponding
probabilities and states of $S$ would also change (cf. Eq.~(\ref{wvfn})).
We will now relate these new probabilities and states with the older ones.

At this stage it is advantageous to express all the probabilities as the
diagonal elements of a density operator (matrix), which is in our case
the reduced density matrix (RDM) of $C$ (denoted by $\rho^C$). This RDM is 
given by $\rho^C = {\rm Tr_{S}}(|\Psi\rangle \langle \Psi|) = gg^{\dagger}$; 
where $g = [g_{i,j}]$ is the matrix representing the tripartite state 
$|\Psi\rangle$ expressed in some product basis states of $C$ and $S$ (cf. 
Eq. (\ref{trprt})).
Using this RDM, the probability associated with a basis state $|\xi\rangle^C$ 
would be $p=~^C\langle \xi| \rho^C |\xi\rangle^C$. This allows us to write the 
new probabilities associated with the new basis states in Eq.~(\ref{newbasis}) 
as,
\begin{eqnarray}
\label{newprbs}
p'_i=p_i+\epsilon k_{ij} ~{\rm and}~ p'_j=p_j-\epsilon k_{ji},
\end{eqnarray}
with $k_{ij}=k_{ji}=~^C\langle \xi_i|\rho^C|\xi_j\rangle^C+~^C\langle \xi_j
|\rho^C| \xi_i\rangle^C$. 

Let us first consider the case when none of the $p_i$ and $p_j$ is zero.
Now if $|\xi'_i \rangle^S$ and $|\xi'_j \rangle^S$ are the new states of 
$S$ corresponding to the two new basis states of $C$ (cf. Eq.~(\ref{wvfn})), 
then in the new scenario, the state $|\Psi\rangle$ can be rewritten as,
\begin{eqnarray}
\label{wvfn1}
|\Psi\rangle = \sqrt{p'_i}|\xi'_i\rangle^C |\xi'_i\rangle^S+\sqrt{p'_j}
|\xi'_j\rangle^C |\xi'_j\rangle^S+\cdots
\end{eqnarray}
Here we only focus on $i$-th and $j$-th states, as other terms are 
unchanged by the considered ET. 
Now using Eqs.~(\ref{newbasis}) and (\ref{newprbs}) in the above 
expression and then comparing the terms associated with the initial basis 
states $|\xi_i\rangle^C$ and $|\xi_j\rangle^C$ from the two different 
expressions of $|\Psi\rangle$ (in Eqs.~(\ref{wvfn}) and~(\ref{wvfn1})), we 
get the following solutions for the new states of $S$:
\begin{eqnarray}
\label{newst1}
|\xi'_i\rangle^S &=& |\xi_i\rangle^S+\epsilon\left(a_{ij}|\xi_i\rangle^S+b_{ij}
|\xi_j\rangle^S\right)~~{\rm and}~\\
\label{newst2}
|\xi'_j\rangle^S &=& |\xi_j\rangle^S-\epsilon\left(a_{ji}|\xi_j\rangle^S+b_{ji}
|\xi_i\rangle^S\right).
\end{eqnarray}
Here $a_{ij}= -\frac{1}{2}k_{ij}p_i^{-1}$ and $b_{ij}= p_j^{1/2}p_i^{-1/2}$
in Eq.~(\ref{newst1}). By interchanging the indices $i$ and $j$ we get
the similar terms in Eq.~(\ref{newst2}).

Let $\rho^{A}(\xi_i)$ and $\rho^{A}(\xi_j)$ be the RDMs of $A$ when $S$ 
is respectively in pure states $|\xi_i\rangle^S$ and $|\xi_j\rangle^S$. 
For example, $\rho^{A}(\xi_i) = 
{\rm Tr_B} (|\xi_i\rangle^S~^S\langle\xi_i|)$. Now if $Q=[Q_{lm}]$ and 
$R=[R_{lm}]$ are the matrices representing respectively the states 
$|\xi_i\rangle^S$ and $|\xi_j\rangle^S$ expressed in some product
basis states of the parts $A$ and $B$, then in terms of these matrices, 
the RDMs of $A$ would be $\rho^{A}(\xi_i)=Q Q^{\dagger}$ and
$\rho^{A}(\xi_j)=R R^{\dagger}$.  
Similarly, the RDMs of $A$ corresponding to the new states, given in
Eqs.~(\ref{newst1}) and~(\ref{newst2}), would be,
\begin{eqnarray}
\label{newrho1}
\rho^{A}(\xi'_i) &=& \rho^{A}(\xi_i)+\epsilon \left(2a_{ij}\rho^{A}(\xi_i)+
2b_{ij}\Delta_{ij}\right)~~{\rm and}~\\
\label{newrho2}
\rho^{A}(\xi'_j) &=& \rho^{A}(\xi_j)-\epsilon \left(2a_{ji}\rho^{A}(\xi_j)+
2b_{ji}\Delta_{ji}\right).
\end{eqnarray}
Here $\Delta_{ij} = \frac{1}{2}(QR^{\dagger}+RQ^{\dagger})$, a Hermitian
matrix. Let us here denote the first order changes in the RDMs in
Eqs.~(\ref{newrho1}) and~(\ref{newrho2})  respectively as 
$\epsilon\rho^{A}_1(ij)$ and $-\epsilon\rho^{A}_1(ji)$; here we have 
\begin{eqnarray}
\label{cngrdm_i}
\rho^{A}_1(ij) = 2a_{ij}\rho^{A}(\xi_i)+ 2b_{ij}\Delta_{ij}\\
\label{cngrdm_j}
\rho^{A}_1(ji)=2a_{ji}\rho^{A}(\xi_j)+2b_{ji}\Delta_{ji}.
\end{eqnarray}
It is worth mentioning that, as the trace (Tr) of any RDM is 1, we must 
have (through Eqs. (\ref{newrho1}) and (\ref{newrho2})),
\begin{eqnarray}
\label{tr_rho1}
{\rm Tr}~ \rho^{A}_1(ij) = {\rm Tr}~\rho^{A}_1(ji)=0.
\end{eqnarray}

Now we can use the following relation (see Appendix A), 
\begin{eqnarray}
\label{entrprltn}
\rho^{A}(\xi'_i)~{\rm log_2} 
\rho^{A}(\xi'_i) &=& \rho^{A}(\xi_i)~{\rm log_2} \rho^{A}(\xi_i) 
+ \epsilon({\rm log_2 e})\rho^{A}_1(ij) \\ \nonumber
& &+ \epsilon \rho^{A}_1(ij)~{\rm log_2} \rho^{A}(\xi_i),
\end{eqnarray}
to obtain the entropy corresponding to the new state $|\xi'_i\rangle^S$. 
Tracing
over both sides of this relation (Eq. (\ref{entrprltn})) and a similar 
relation for the state $|\xi'_j\rangle^S$, we obtain respectively the 
following entropies for the new states of $S$,
\begin{eqnarray}
\label{new_entr1}
\mathcal{S}'_i&=&\mathcal{S}_i-\epsilon{\rm Tr}~\rho^{A}_1(ij)~{\rm log_2~} 
\rho^{A}(\xi_i)~~{\rm and}~\\
\label{new_entr2}
\mathcal{S}'_j&=&\mathcal{S}_j+\epsilon{\rm Tr}~\rho^{A}_1(ji)~{\rm log_2~} 
\rho^{A}(\xi_j).
\end{eqnarray}
Here we used Eq.~(\ref{tr_rho1}) to get these relations. Let us now denote the
first order changes in entropies in Eqs.~(\ref{new_entr1}) and~(\ref{new_entr2}) 
as $-\epsilon\mathcal{S}^1_{ij}$ and $\epsilon\mathcal{S}^1_{ji}$ respectively, 
with, 
\begin{eqnarray}
\label{cngentrp_i}
\mathcal{S}^1_{ij} = {\rm Tr}~\rho^{A}_1(ij)~{\rm log_2~} \rho^{A}(\xi_i)\\
\label{cngentrp_j}
\mathcal{S}^1_{ji} = {\rm Tr}~\rho^{A}_1(ji)~{\rm log_2~} \rho^{A}(\xi_j).
\end{eqnarray}
Now using these new entropies in Eqs. (\ref{new_entr1}) and (\ref{new_entr2})
along with the new probabilities in 
Eq.~(\ref{newprbs}), we get the new average entropy (cf. Eq. (\ref{aventr})):
\begin{eqnarray}
\label{newaventr}
\bar{\mathcal{S}}'=\sum\nolimits_{l=1}^{D} p'_l \mathcal{S}'_l
=\bar{\mathcal{S}}+\epsilon \bar{\mathcal{S}}_1, 
\end{eqnarray}
where,
$\bar{\mathcal{S}}_1=k_{ij}\mathcal{S}_i-p_i\mathcal{S}^1_{ij}-
k_{ji}\mathcal{S}_j+p_j\mathcal{S}^1_{ji}$. This $\bar{\mathcal{S}}_1$ is the 
first order change in the average entropy due to an ET of the $i$th and $j$th 
basis states.

Before we set the optimality condition, let us now check the special cases 
when both $p_i$ 
and $p_j$ are or one of them is zero. When $p_i=p_j=0$, then $k_{ij}=k_{ji}=0$.
Therefore, $p'_i=p'_j=0$ (see Eq. (\ref{newprbs})). Which
implies that $\bar{\mathcal{S}}_1$ is zero. On the other hand, when $p_i\neq0$
and $p_j=0$, we gave again $k_{ij}=k_{ji}=0$. From Eq. (\ref{newprbs}) we have
$p'_i = p_i$ and $p'_j=0$. From Eq. (\ref{newst1}), we also have $|\xi'_i\rangle^S 
= |\xi_i\rangle^S$. This implies that, $\bar{\mathcal{S}}_1 = 0$. 
In these two special cases the first order change in average entropy due to 
any elementary transformation (ET) of the two basis states concerned is always
zero; therefore we need not consider these cases to determine whether some basis 
set of measurement is optimal.  

So, the desired {\it optimality condition} can be obtained by equating to zero 
the first 
order change in the average entropy due to an ET of any two basis states for which 
corresponding probabilities are nonzero. This condition is given by the following 
equation $\bar{\mathcal{S}}_1=0$ (cf. Eq. (\ref{newaventr})) or,
\begin{eqnarray}
\label{cndtn1}
k_{ij}\mathcal{S}_i-p_i\mathcal{S}^1_{ij}=k_{ji}\mathcal{S}_j-
p_j\mathcal{S}^1_{ji},
\end{eqnarray}
for all $i$ and $j$ for which corresponding probabilities are
nonzero. 

\section{Study of W and GHZ states}
\noindent There are two types of genuine pure tripartite states (which cannot be 
converted to each other by the SLOCC operation \cite{dur00}), namely $W$ and $GHZ$
states. These states are defined as follows,
\begin{eqnarray}
\label{wstate}
|W\rangle = \frac{1}{\sqrt{3}}(|100\rangle + |010\rangle +
|001\rangle)\\
\label{ghzstates}
|GHZ\rangle = \frac{1}{\sqrt{2}}(|000\rangle + 
|111\rangle).
\end{eqnarray}
Main difference between these two states is, the entanglement in $|W\rangle$ 
is robust in the sense that when one qubit is traced out other two qubits 
remain entangled, on the contrary, the entanglement in $|GHZ\rangle$ is 
fragile due to the fact that when one qubit is traced out other two qubits  
become unentangled. In what follows, we will study the average entanglement 
($\bar{\mathcal{S}}$) between two qubits for both the tripartite states. In 
particular we show that, for $|W\rangle$ state, $0 < E_F < E_A < 1$, on the 
contrary for $|GHZ\rangle$ state, $E_F = 0$ and  $E_A = 1$, i.e. those two 
quantities attain their extreme possible values for $|GHZ\rangle$ state. 
(The quantities $E_F$ and $E_A$ are defined in Eqs. (\ref{entfrm1}) 
and (\ref{entasst1}) respectively.)    

First thing we will see here is that, for $|W\rangle$ and 
$|GHZ\rangle$ states, if the measurement basis set is the set of eigenstates 
of the RDM of $C$ (i.e., $\rho^C$), then the corresponding average entropy 
($\bar{\mathcal{S}}$) associated with the decomposition of $\rho^{AB}$ 
($\equiv \rho^S$) is optimal. In other words, for those two states, the 
eigenstates of $\rho^C$ satisfy the optimality condition. 

We note that when the measurement is performed by the eigenstates of 
$\rho^{C}$, then $k_{ij}=k_{ji}=0$. This reduces Eq.~(\ref{cndtn1}) to
\begin{eqnarray}
\label{1cndtn_rdm}
 {\rm Tr}~\Delta_{ij}~
{\rm log_2~} \rho^{A}(\underline{\xi}_i)={\rm Tr}~\Delta_{ji}~
{\rm log_2~} \rho^{A}(\underline{\xi}_j),
\end{eqnarray}
where $\{\underline{\xi}\}$ is the set of the eigenstates of $\rho^{S}$.
The easy way to check our claim is to do the Schmidt 
decomposition of the states, as this decomposition is constructed by 
the eigenstates of both $\rho^C$ and $\rho^S$.

For the state $|W\rangle$, the Schmidt decomposition 
takes the following form: $|W\rangle = \sqrt{\frac{2}{3}}
[\frac{1}{\sqrt{2}}(|10\rangle^S + |01\rangle^S)] |0\rangle^C + 
\frac{1}{\sqrt{3}} |00\rangle^S |1\rangle^C$. Here 
$|\xi_1\rangle^S = \frac{1}{\sqrt{2}}(|10\rangle^S + |01\rangle^S)$ 
and $|\xi_2\rangle^S = |00\rangle^S$ are the eigenstates of 
$\rho^S$ corresponding to the non-zero eigenvalues, and 
$|0\rangle^C$ and $|1\rangle^C$ are the eigenstates of $\rho^C$. 
The $2\times 2$ matrices representing $|\xi_1\rangle^S$ and 
$|\xi_2\rangle^S$ are respectively, $P = \frac{1}{\sqrt{2}} 
\bigl(\begin{smallmatrix} 0&1\\ 1&0 \end{smallmatrix} \bigr)$ and 
$Q = \bigl(\begin{smallmatrix} 1&0\\ 0&0 \end{smallmatrix} \bigr)$.
From these matrices we get, $\rho^{A}(\underline{\xi}_1) = 
PP^{\dagger} = \frac12 
\bigl(\begin{smallmatrix} 1&0\\ 0&1 \end{smallmatrix} \bigr)$ and 
$\rho^{A}(\underline{\xi}_2) = QQ^{\dagger} = 
\bigl(\begin{smallmatrix} 1&0\\ 0&0 \end{smallmatrix} \bigr)$. We 
also get $\Delta_{12} = \Delta_{21} = 
\frac12 (PQ^{\dagger} + QP^{\dagger}) = \frac{1}{2\sqrt{2}} 
\bigl(\begin{smallmatrix} 0&1\\ 1&0 \end{smallmatrix} \bigr)$. By 
simple matrix manipulations, we now can check that these matrices 
actually satisfy the optimality condition given in 
Eq. (\ref{1cndtn_rdm}). This shows that if we do measurement in the 
basis set $\{|0\rangle^C, |1\rangle^C\}$, we will get an optimal 
value of the average entanglement ($\bar{\mathcal{S}}$). To get 
this optimal value, we note that the entropies of the states 
$|\xi_1\rangle^S$ and $|\xi_2\rangle^S$ are 1 and 0 respectively, 
and they appear in the decomposition with corresponding probabilities 
2/3 and 1/3. Therefore the optimal value of the average entropy 
($\bar{\mathcal{S}}$) in this case is 2/3.

For the state $|GHZ\rangle$, the Schmidt decomposition
takes the following form: $|GHZ\rangle = 
\frac{1}{\sqrt{2}}|00\rangle^S |0\rangle^C + 
\frac{1}{\sqrt{2}}|11\rangle^S |1\rangle^C$. Here
$|\xi_1\rangle^S = |00\rangle^S$ and $|\xi_2\rangle^S = |11\rangle^S$ 
are the eigenstates of $\rho^S$ corresponding to the non-zero eigenvalues, 
and $|0\rangle^C$ and $|1\rangle^C$ are the eigenstates of $\rho^C$. A 
similar calculation as above shows that the measurement basis set 
$\{|0\rangle^C, |1\rangle^C\}$ is optimal and corresponding optimal 
value of $\bar{\mathcal{S}}$ is 0.

Though the optimality test can tell us whether a basis set of measurement is 
optimal, it cannot tell us whether the corresponding average entropy is a 
maximum (i.e., $E_A$) or minimum (i.e., $E_F$). 
Fortunately, for any $2\times2\times2$ pure tripartite 
state, it is possible to calculate $E_F$ and $E_A$ (cf. Eqs. (\ref{entfrm1}) 
and (\ref{entasst1})) exactly. Now we will do the detail exact calculation 
for both the states ($|W\rangle$ and $|GHZ\rangle$) to verify the above 
results as well as to find $E_F$ and $E_A$ for those two states.

\begin{figure} []
\begin{center}
\includegraphics[width=10.0cm]{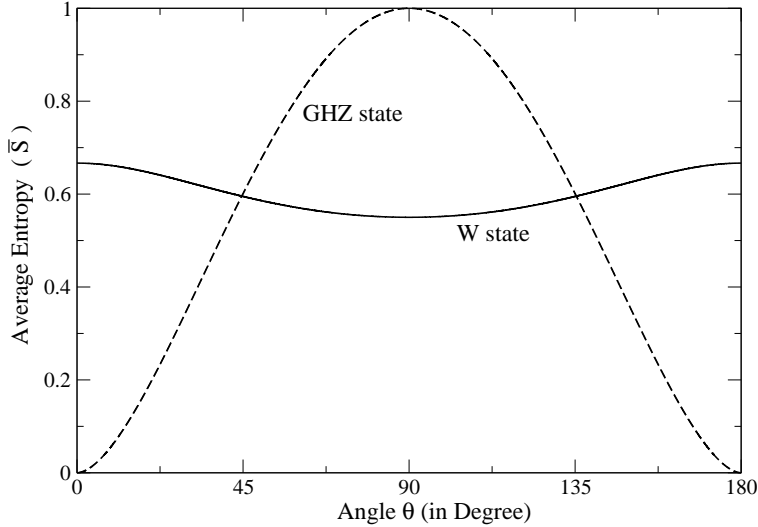}
\end{center}
\caption{The average entropy ($\bar{\mathcal{S}}$) as a function 
of rotation angle $\theta$.}
\label{wghz}
\end{figure}

To proceed, let us take the most general set of measurement basis as 
$\{\rm cos \frac{\theta}{2} |0\rangle + e^{i\phi} sin \frac{\theta}
{2}|1\rangle,~sin \frac{\theta}{2} |0\rangle - e^{i\phi} cos \frac{\theta}
{2}|1\rangle \}$ (two diagonally opposite points on {\it Bloch sphere}), where 
$0\le \theta \le \pi$ and $0\le \phi < 2\pi$. If we now denote this basis set 
by $\{|\xi_1\rangle^C, |\xi_2\rangle^C \}$, then the $|W\rangle$ state can be 
rewritten as, 

\begin{eqnarray}
|W\rangle &=& \frac{1}{\sqrt{3}}\sqrt{|a|^2+2|b|^2}~|\xi_1\rangle^C~(a^\prime
|00\rangle^S + b^\prime |10\rangle^S + b^\prime |01\rangle^S) \nonumber \\
&&+\frac{1}{\sqrt{3}}\sqrt{|c|^2+2|d|^2}~|\xi_2\rangle^C~(c^\prime
|00\rangle^S + d^\prime |10\rangle^S + d^\prime |01\rangle^S)
\label{wstate1}
\end{eqnarray}
Here $a^\prime = \frac{a}{\sqrt{|a|^2+2|b|^2}}$, 
$b^\prime = \frac{b}{\sqrt{|a|^2+2|b|^2}}$, 
$c^\prime = \frac{c}{\sqrt{|c|^2+2|d|^2}}$ and 
$d^\prime = \frac{d}{\sqrt{|c|^2+2|d|^2}}$. In terms of $\theta$ and $\phi$, 
these parameters are given by ${\rm a = e^{-i\phi} sin\frac{\theta}{2}}$, 
${\rm b = cos\frac{\theta}{2}}$, ${\rm c =-e^{-i\phi} cos\frac{\theta}{2}}$ 
and ${\rm d = sin\frac{\theta}{2}}$. The operational interpretation of the 
above expression of the $|W\rangle$ state (Eq.~(\ref{wstate1})) is that if 
one does quantum measurement on $C$ by the 
basis set $\{|\xi_1\rangle^C, |\xi_2\rangle^C \}$, then $S$ will be found in 
the state $|\xi_1\rangle^S$ with probability $p_1 = 
\frac{1}{3}(|a|^2+2|b|^2)$ and in the state $|\xi_2\rangle^S$ with 
probability $p_2 = \frac{1}{3}(|c|^2+2|d|^2)$; here $|\xi_1\rangle^S = a^\prime
|00\rangle^S + b^\prime |10\rangle^S + b^\prime |01\rangle^S$ and $|\xi_2\rangle^S 
= c^\prime |00\rangle^S + d^\prime |10\rangle^S + d^\prime |01\rangle^S$. It is 
easy to find $2\times 2$ RDMs of $A$ (i.e., $\rho^A$) for both the states 
$|\xi_1\rangle^S$ 
and $|\xi_2\rangle^S$. From these RDMs, one can get corresponding entropies and 
the average entropy thereafter (see Appendix B). This average entropy 
($\bar{\mathcal{S}}$) is only function of $\theta$; $\phi$ does not appear in its 
expression. The minima and maxima can be found from $\frac{\partial 
\bar{\mathcal{S}}}{\partial \theta} = 0$. We find that global maximum corresponds 
to $\theta = 0$ (or equivalently $\theta = \pi$) and global minimum corresponds to 
$\theta = \frac{\pi}{2}$. These can be seen clearly from Fig.~\ref{wghz}. We may 
note that, when $\theta$ = 0 or $\pi$, within a global phase factor, the 
measurement basis set becomes $\{|\xi_1\rangle^C, |\xi_2\rangle^C \} 
\equiv \{|1\rangle, |0\rangle\}$. This asserts that the 
eigenstates of $\rho^C$ are the optimal basis set of measurement. For the 
$|W\rangle$ state, the maximum and minimum possible values of $\bar{\mathcal{S}}$ 
are 2/3 ($\simeq$ 0.67) and 
$log_23 - \frac{\sqrt{5}}{3} log_2 (\frac{3+\sqrt{5}}{2})$ ($\simeq$ 0.55) 
respectively. So for this tripartite state $E_F \simeq 0.55$ and
$E_A \simeq 0.67$.

A similar calculation for the $|GHZ\rangle$ state is also done (see Appendix C); 
the global 
minimum and maximum occur at $\theta = 0$ (or equivalently at $\theta = \pi$) and 
at $\theta = \frac{\pi}{2}$ respectively. This minimum and maximum values of the 
average entropy ($\bar{\mathcal{S}}$) are respectively 0 and 1 (see Fig.~\ref{wghz}).
So for this tripartite state $E_F = 0$ and $E_A = 1$.

\section{Conclusion}
\noindent
Any mixed bipartite state can be decomposed in innumerable ways. By the
Hughston-Jozsa-Wootters theorem, any finite decomposition 
of the mixed state can be seen 
as a result of some local measurement on a separate part added in a pure 
state with the bipartite system. Since each of these decompositions can be 
assigned an average entanglement, by choosing an appropriate measurement 
basis set we will be able to get an optimal value of the average 
quantity. For a given tripartite system in some pure state, we have 
derived here an optimality condition for the measurement basis set; when 
satisfied, the average entanglement associated with the decomposition due 
to the measurement will be optimal. By the use of {\it ancilla}, 
this optimality condition would also be helpful in finding the 
entanglement of formation and the entanglement of assistance for a given 
bipartite mixed state (without any reference to tripartite state).

In the second part, we have studied two inequivalent tripartite states 
($W$ and $GHZ$); in particular, we 
showed that the set of eigenvectors of the RDM of a qubit is an optimal 
basis set of measurement (for both the states). 
To verify the result and to find the maximum and minimum possible average 
bipartite entanglement ($E_A$ and $E_F$ respectively), we have also done 
the detailed exact calculations for both the states. The values of $E_A$ 
and $E_F$ found for the states show why they are inequivalent.

\section*{Acknowledgements}
I thank Prof. S. Ramasesha (SR) and Prof. Diptiman Sen for useful discussions. 
I also acknowledge SR's financial support through his various projects from 
IFCPAR and DST, India.

\vspace*{0.37truein}
\centerline{\bf Appendix A}
\vspace*{0.14truein}
\noindent 
The expansion of the operator $\rho^{A}(\xi'_i)~{\rm log_2} \rho^{A}(\xi'_i)$ 
in the power series of the parameter $\epsilon$ is difficult as 
$\rho^{A}_1(ij)$ and $\rho^{A}(\xi_i)$ do not commute in general. 
Ultimately since we only need to know the trace of this operator (to find 
entropy), and as ${\rm Tr~ MN = Tr~NM}$ for any two compatible 
finite matrices $M$ and $N$, we can proceed in the following way to get 
the first order change in the operator. 

If $\lambda_l$ is the $l$-th eigenvalue of $\rho^{A}(\xi_i)$
and $\lambda^1_l$ is the expectation value of $\rho^{A}_1(ij)$ in the
corresponding eigenstate of $\rho^{A}(\xi_i)$, 
then the expectation value of the operator $\rho^{A}(\xi'_i)~
log_2~ \rho^{A}(\xi'_i)$ in the eigenstate would be
$(\lambda_l+\epsilon \lambda^1_l)~log_2~\lambda_l(1+\epsilon 
\frac{\lambda^1_l}{\lambda_l})$. This equals to $\lambda_l~log_2
\lambda_l + \epsilon (log_2 e)~\lambda^1_l + \epsilon \lambda^1_l 
~log_2 \lambda_l$, using $log_2(1+\epsilon \frac{\lambda^1_l}
{\lambda_l}) = \epsilon (log_2 e)~\frac{\lambda^1_l}{\lambda_l}$. 
This eventually
suggests the operator relation we use in the text. This operator 
relation is not ill-defined due to the last term since when 
$\lambda_l$ is zero, $\lambda^1_l$ also becomes zero (this can
be understood by singular value decomposition of the matrix $Q$; 
see form of $\rho^{A}$ and $\rho^{A}_1$ in the main text). 

\vspace*{0.3truein}
\centerline{\bf Appendix B}
\vspace*{0.14truein}
\noindent 
For the $W$ state, when $S$ (=$A$+$B$) is in the bipartite 
state  $|\xi_1\rangle^S = a^\prime |00\rangle^S + b^\prime |10\rangle^S 
+ b^\prime |01\rangle^S$, the RDM for the part $A$ is given by,\\ 

\begin{center}
$\rho^A(\xi_1) = \left( \begin{array}{cc} 
|a'|^2+|b'|^2 & a'(b')^* \\
b'(a')^* & |b'|^2 \end{array} \right),$
\end{center}

\noindent with $\lambda^\pm_1 = \frac{1}{2}\left[1\pm 
\frac{sin \frac{\theta}{2}}{1+cos^2 \frac{\theta}{2}}\sqrt{1+
3cos^2\frac{\theta}{2}}\right]$ being its
two eigenvalues expressed in terms of the parameter $\theta$. The entropy of 
this state is given by $\mathcal{S}_1 =  -\lambda^+_1 log_2 \lambda^+_1 - 
\lambda^-_1 log_2 \lambda^-_1$. Similarly, in case of the state 
$|\xi_2\rangle^S = c^\prime |00\rangle^S + d^\prime |10\rangle^S + 
d^\prime |01\rangle^S$, the entropy is given by $\mathcal{S}_2 = 
-\lambda^+_2 log_2 \lambda^+_2 - \lambda^-_2 log_2 \lambda^-_2$, where
$ \lambda^\pm_2 = \frac{1}{2}\left[1\pm 
\frac{cos \frac{\theta}{2}}{1+sin^2 \frac{\theta}{2}}\sqrt{1+
3sin^2\frac{\theta}{2}}\right]$. These
two states come with probabilities 
$p_1 = \frac{1}{3}(1 + cos^2\frac{\theta}{2})$ and 
$p_2 = \frac{1}{3}(1 + sin^2\frac{\theta}{2})$ respectively when quantum 
measurement is done on part $C$ by the basis set 
$\{|\xi_1\rangle^C, |\xi_2\rangle^C \}$. The average entropy  
localized in $S$ due to measurement on $C$ is therefore, $\bar{\mathcal{S}} 
= p_1 \mathcal{S}_1 + p_2 \mathcal{S}_2$. We may here note that, 
$\bar{\mathcal{S}}$ depends only on the parameter $\theta$.

\vspace*{0.3truein}
\centerline{\bf Appendix C}
\vspace*{0.14truein}
\noindent 
For the $GHZ$ state, when a measurement is done on $C$ by the basis set 
$\{|\xi_1\rangle^C, |\xi_2\rangle^C \}$, $S$ will be found in the bipartite 
state $|\xi_1\rangle^S = a |00\rangle^S + b |11\rangle^S$ with probability 
$p_1 = \frac 12$ and in the state $|\xi_2\rangle^S = c |00\rangle^S + 
d |11\rangle^S$ with probability $p_2 = \frac 12$. Here, 
$a = cos \frac{\theta}{2}$, $b = e^{-i\phi} sin  \frac{\theta}{2}$, 
$c = sin \frac{\theta}{2}$ and 
$d = -e^{-i\phi} cos  \frac{\theta}{2}$. When $S$ is in $|\xi_1\rangle^S$, 
the RDM for the part $A$ is given by, \\

\begin{center}
$\rho^A(\xi_1) = \left( \begin{array}{cc} 
|a|^2 & 0 \\
0 & |b|^2 \end{array} \right),$
\end{center}

\noindent with eigenvalues $\lambda_{11} = cos^2\frac{\theta}{2}$ and 
$\lambda_{12} = \sin^2\frac{\theta}{2}$ expressed in terms of the 
parameters $\theta$. The entropy of this state is given by $\mathcal{S}_1 = 
-\lambda_{11} log_2 \lambda_{11} -\lambda_{12} log_2 \lambda_{12}$. 
Similarly, the entropy for the state $|\xi_2\rangle^S$ is given by 
$\mathcal{S}_2 = 
-\lambda_{21} log_2 \lambda_{21} -\lambda_{22} log_2 \lambda_{22}$; where,
$\lambda_{21} = \sin^2\frac{\theta}{2}$ and 
$\lambda_{22} = \cos^2\frac{\theta}{2}$. The average entropy
localized in $S$ due to measurement on $C$ is therefore, $\bar{\mathcal{S}} 
= p_1 \mathcal{S}_1 + p_2 \mathcal{S}_2$ or $- (cos^2\frac{\theta}{2}~ 
log_2~cos^2\frac{\theta}{2}$ + 
$sin^2\frac{\theta}{2}~log_2~sin^2\frac{\theta}{2})$.

\end{document}